\documentclass[12pt]{article}

\usepackage{bm}
\usepackage{amssymb}
\usepackage{enumerate}

\usepackage{graphicx} 
\begin{document}
\title{Strictly solvable $d=4$  quantum field theory models with unitarity violation.
  Possible implications for  baryon asymmetry of the Universe  and dark matter } 


\author{N.V.Krasnikov \thanks{ NikolayKrasnikov6@gmail.com} \\
\\Institute for Nuclear Research RAS\\
\\ Moscow, Russia
\\and
\\ JINR Dubna, Russia
}
\maketitle
\begin{abstract}

  In this paper we consider 
  quantum field theory with unitarity violation. We construct several $d = 4$ strictly solvable models with
  unitarity violation.
  As a possible application of the   models with unitarity violation we
   construct two models explaining
  the observed baryon asymmetry of the Universe and the dark matter existence.
  In considered  models dark matter particles have nonzero  baryon number and 
  the total baryon number of the Universe being the sum of the baryon number of the matter and dark matter  is equal to zero.
  Dark matter paricles have mass  $\sim 6~GeV$. Proposed models predict negligible cross sections of the
  dark matter with nuclei and electrons. It means that underground experiments will fail to discover dark matter.

\end{abstract}


 \newpage

\section{INTRODUCTION}

As is well known the general properties of the S-matrix are  the following  \cite{Bog}:      

1. The invariance of the S-matrix under Poincare group transformations.

2. The stability of the vacuum and one particle state.

3. The unitarity of the S-matrix.

4. The locality and microcausality of the S-matrix.

The property of the unitarity is a  direct consequence of the hermiticity of the Hamiltonian.
The hermiticity of the Hamiltonian means the conservation of the probabilty. Nevertheless
nonhermitean Hamiltonians are often used for the description of the evolution of unstable particles  and
  non-closed systems. The typical example is the description of the
kaon oscillations on the base of nonhermitean Hamiltonian  \cite{ParticleData}. It is assumed that the generalizations of kaon system including
the products of  kaon decays is a system which is described by the hermitean Hamiltonian and unitary
$S$-matrix. The introduction  of Pauli-Villars regularization \cite{PAULI} and
its generalization to the case of nonabelian gauge theories \cite{Slavnov}    allows to get rid of
ultraviolet divergences in Feynman integrals by the introduction of negative norm states in the spectrum of the model.
In quantum gravity the introduction of additional terms to Einstein gravity  with higher order
derivatives makes quantum gravity renormalizable theory  \cite{Stelle}.
The use of  Pauli-Villars regularization and its generalizations leads to violation of the
unitarity of the $S$-matrix \cite{Leon,Leon1,Kubo,Krasnikov}.
For instance, for the $\phi^4$-model Pauli-Villars regularization
  can be reformulated  by the introduction  of additional massive field at the level
  of free Lagrangian and the use of  nonhermitean interaction Lagrangian. However
       there is no any reasonable physical interpretation of the models with
  Pauli-Villars regularization.
 Therefore it is important  to study the examples of the nonunitary quantun field theory  models    to understand the level of
pathology of such systems. Also it is interesting to find  additional possibilities  arising
with the nonhermitean Lagrangians(Hamiltonians).

In this
paper we study
  quantum field theory with unitarity violation. We construct several strictly solvable models with
  unitarity violation.
  As a possible application of  the models with unitarity violation we  propose two models   explaining
   the observed baryon asymmetry of the Universe and dark matter existence.
   In consisered  models dark matter particles have nonzero  baryon number.
   The total baryon number of the Universe being the sum of the baryon number of the matter and dark matter  is equal to zero.
   The mass of dark matter particles is $\sim 6~GeV$.
     Proposed models predict negligible cross sections of the
  dark matter with nuclei and electrons. It means that underground experiments will fail to discover dark matter.

    Models with unitarity violation also   violate $CPT$ theorem,
   in particular, the decay width of the particle is  not equal to the decay width of the untiparticle.
   Therefore       the     search for  possible nonunitarity of the $S$-matrix in K-meson decays
   is especially interesting.

The organization of the paper is the following. In the next section we give examples of the nonhermitean interaction
Lagrangians which lead to
 strictly  solvable models.
In section 3 we propose two models with unitarity violation which allow to explain
 the baryon asymmetry of the Universe and dark matter. 
   Section 4  contains concluding remarks.
 

  \section{EXACTLY SOLVABLE FOUR DIMENSIONAL MODELS WITH UNITARITY VIOLATION}

  Consider the  model with  complex scalar field and with the Lagrangian
  \begin{equation}
    L = L_0 + L_{int} \,.
    \label {Lag}
    \end{equation}
   Free Lagrangian $L_o$ has standard  form  \cite{Bog}
\begin{equation}
  L_0 = \partial^{\mu}\phi(x) \partial_{\mu}\phi^{*}(x) - m^2\phi^{*}(x)\phi(x) \,.
\label{freelag}
\end{equation}
For the Lagrangian (\ref{freelag}) the solution for the quantized field $\phi(x)$ can be written in the  form \cite{Bog}
\begin{equation}
\phi(x) = \int d^4p [ \exp(-ipx)\frac{\hat{a}(\vec{p})}{2p^0(2\pi)^3} +\exp(ipx)\frac{\hat{b}^+(\vec{p})}{2p^0(2\pi)^3}] \,,
\label{sol}
\end{equation}
where $ [\hat{a}(\vec{p}), \hat{a}^+(\vec{p}^{'})] =  [\hat{b}(\vec{p}), \hat{b}^+(\vec{p}^{'} )] = 2p^0(2\pi)^3 \delta(\vec{p} - \vec{p}{'} ) $
and $p^{0} = \sqrt{\vec{p}^2 +m^2}$.
As a simplest example of the model with  unitarity violation  we take the interaction Lagrangian $L_{int}$ in the form
\begin{equation}
  L_{int} = \frac{\lambda}{4!}\phi^4(x) \,.
  \label{Lint1}
\end{equation}
The Lagrangian (\ref{Lint1}) is not hermitean, as a consequence the property of the unitarity is lost.
Due to the fact that the free propagators $<0|T(\phi(x), \phi(y)|0>$ and  $<0|T(\phi^{*}(x), \phi^{*}(y)|0>$
are equal to zero all quantum corrections to the interaction Lagrangian (\ref{Lint1}) vanish and the effective action
coincides with free action
\begin{equation}
  S_{eff} = \int d^4x  [L_0 + L_{int}] \,.
\label{Seff}
\end{equation}
As a consequence all quantum corrections to the $S$-matrix vanish and  only the reaction
\begin{equation}
  \phi(\vec{p}_1) ~+~ \phi(\vec{p}_2) \rightarrow  \phi^+(\vec{p}_3) ~+~ \phi^+(\vec{p}_4) \, 
      \label{reaction}
\end{equation}
with the $T$-matrix
\begin{equation}
  T( \phi(\vec{p}_1) ~+~ \phi(\vec{p}_2) \rightarrow  \phi^+(\vec{p}_3) ~+~ \phi^+(\vec{p}_4)) =\frac{ \lambda}{2} \,
  \label{Tmatrix}
\end{equation}
is possible.
Here   $ \phi(\vec{p}) =   a^{+}(\vec{p})|0> $,  $ \phi^{+}(\vec{p}) = b^{+}(\vec{p})|0> $ and 
$   <  \phi^+(\vec{p}_4) \phi^+(\vec{p}_3) |(\hat{S}-1)|\phi(\vec{p}_2) \phi(\vec{p}_1)  > = i (2\pi)^4
T( \phi(\vec{p}_1) ~+~ \phi(\vec{p}_2) \rightarrow  \phi^+(\vec{p}_3) ~+~ \phi^+(\vec{p}_4))\delta^{(4)}(p_1 + p_2 - p_3 - p_4)$.
For the considered model the well known formula for the $S$-matrix
  $ S = T(exp(i\int L_{int}(x)d^4x))$ is reduced to
 $  S = exp(i\frac{\lambda}{4!}\int d^4x \phi^4(x))$.
As a consequence we  find that the $S$-matrix is local and microcausal according to the definition of ref.(\cite{Bog}).
Note that other  reactions like 
\begin{equation}
  \phi^{+}(\vec{p}_1) ~+~ \phi^{+}(\vec{p}_2) \rightarrow  \phi(\vec{p}_3) ~+~ \phi(\vec{p}_4) \, 
      \label{reactionop}
\end{equation}
have  zero cross sections.
It is interesting to mention that classical  equations  for the Lagrangian (\ref{Lag})
\begin{equation}
  ( \partial^{\mu}\partial_{\mu} + m^2)\phi^*(x) = \frac{\lambda}{3!}\phi^3(x) \,,
  \label{eq1}
  \end{equation}
\begin{equation}
  ( \partial^{\mu}\partial_{\mu} + m^2)\phi(x) = 0 \,
  \label{eq2}
  \end{equation}
have the single solution $\phi(x) = \phi^*(x) = 0$.
Consider now a model with  two complex scalar fields $\chi(x)$, $\phi(x)$ and  with the
interaction Lagrangian
\begin{equation}
  L_{int} = \frac{\lambda}{2}\chi(x)\phi^2(x)  \,.
  \label{model2}
  \end{equation}
  For this model 
   the  action is invariant under $CPT$ transformation
 $\phi(x) \rightarrow \phi_{CPT}(x) = \phi(-x)$, $\chi(x) \rightarrow \chi_{CPT}(x) = \chi(-x)$.
 However  the CPT theorem prediction\footnote{See for example \cite{Bog}.}
that total decay widths of particle and antiparticle coincide 
is not valid for the models with unitarity violation because the CPT theorem is based on the aditional  assumption that the Lagrangian is hermitean.
Really as in a previous example radiative corrections vanish for the interaction ({\ref{model2})  
   and for $m_{\chi} > 2 m_{\phi}$ the $\chi$-particle decays into two $\phi^*$ particles with the decay width
  $\Gamma(\chi \rightarrow \phi^* +\phi^*) =  \frac{\lambda^2}{32\pi} \sqrt{1 - \frac{4m^2_{\phi}}{m^2_{\chi}}}$
      while the decay width of the antiparticle $\chi^*  \rightarrow \phi + \phi $ is equal to zero and the $\chi^*$-antiparticle is stable.

      One can find that in general case quantum corrections  to  the model with the complex scalar field $\phi(x)$ and with the
      interaction Lagrangian
\begin{equation}
  L_{int,gen} = \sum_{k=3}^{\infty}       \int C_k(x; x_1...x_k) \phi(x_1)...\phi(x_k)d^4x_1...d^4x_k  \,
  \label{intaction}
\end{equation}
vanish and only the reactions $\phi(\vec{p}_1) ~+~ \phi(\vec{p}_2)+ ..\phi(\vec{p}_l)     \rightarrow
\phi^{+}(\vec{p}_{l+1}) ~+~ \phi^{+}(\vec{p}_{l+2}) + ... \phi^{+}_{l+n})$ are possible. 
As another  example consider model with neutral scalar field $\phi(x)$ and complex scalar field $\Phi(x)$
with the Lagrangian
\begin{equation}
  L = L_{0\phi} +  L_{0\Phi} + L_{int} \,,
  \label{Ltot1}
\end{equation}
\begin{equation}
 L_{0\phi} =\frac{1}{2}[ \partial^{\mu}\phi(x) \partial_{\mu}\phi(x) - m^2\phi(x)\phi(x)] \,,
 \label{Lphi}
\end{equation}
\begin{equation}
 L_{0 \Phi}=     \partial^{\mu}\Phi^{*}(x) \partial_{\mu}\Phi(x) - m^2\Phi^{*}(x)\Phi(x) \,,
 \label{LPhi}
 \end{equation}
\begin{equation}
  L_{int}  = -\lambda(\phi(x) +\Phi(x))^4 \,,
  \label{Lint}
  \end{equation}
One can find that for the model    with the interaction Lagrangian  (\ref{Lint}) quantum corrections to the effective action
depend on  $\phi_1(x) = \phi(x) + \Phi(x) $ field. The effective action does not contain   the  $\Phi^{*}(x)$ field.

The generalization of the previous examples  to the case of fermion fields is straightforward. Namely,
as a simplest example consider two fermion Dirac  fields  $\psi_a(x)$, $\psi_b(x)$
with the interaction Lagrangian
\begin{equation}
  L_{I} = G\bar{\psi}_a(x)\psi_b(x)\bar{\psi}_a(x)\psi_b(x) \,.
  \label{Lferm}
\end{equation}
One can find that as in the scalar case for  the interaction Lagrangian (\ref{Lferm}) quantum corrections to the effective action
also vanish. The generalization of the Lagrangian (\ref{Lferm}) to the more general form as
the Lagrangian (\ref{intaction}) is straightforward.

Also  the following  generalization of the previous examples   is
very interesting. Namely, consider the  complex scalar field $\phi(x)$ which interacts with the standard model fields as
\begin{equation}
  L_{\phi,SM} = \lambda \phi(x)V_{SM}(H(x),q(x),l(x)) \,,
  \label{lagSMphi}
  \end{equation}
  where $V_{SM}(H(x),q(x),l(x))$ is a function of the Higgs boson, quark and lepton fields. For example
  $V_{SM}(H(x),q(x),l(x)) = H^{+}(x)H(x)$ or  $V_{SM}(H(x),q(x),l(x)) = \bar{q}(x)q(x)$. Here $H(x) = (H_1(x), H_2(x))$ is the Higgs isodoublet field. 
  One can find that for the Lagrangian (\ref{lagSMphi}) the radiative corrections for the $\phi^*(x)$ field
  vanish while the radiative corrections for the SM fields are different from zero and the effective action has the form
  \begin{equation}
    S_{eff} = \sum_{k =1}^{\infty}\int \prod_{l=1}^{l=k} d^4x_l G_k(H, q, l)\phi(x_1)...\phi(x_k) \,.
    \label{Seffgen}
  \end{equation}
  The terms  $\phi...\phi\phi^{*}...\phi^{*} G_k(H, q, l)  $ are absent in the effective action.

\section{Models with unitarity violation explaining the baryon asymmetry of the Universe}

As a possible application of the considered models with unitarity violation we propose to use such models
for an explanation of the baryon asymmetry of the Universe \cite{Rubakov} - \cite{Zurek} and the existence of  dark matter\footnote{ 
  As a review
  of the baryogenesis and dark matter problems see for example \cite{Rubakov, Kolb, B5}.
  Note also  that there are models relating the dark matter existence and the baryon
  asymmetry violation, as a review see \cite{Zurek}
}.
As a simplest realization of the idea with the use of the models with unitarity violation
we consider the four fermion interaction 
 of colourless electrically neutral $SU_c(3) \otimes SU_L(2) \otimes U(1)$ singlet
fermion $\psi(x) $ with quarks . The interaction of the $\psi$ field with
colour quarks $u^{\alpha}_{k} = (u^{\alpha}, c^{\alpha}, t^{\alpha})$ and
$d^{\alpha}_{k} = (d^{\alpha}, s^{\alpha}, b^{\alpha})$  has the form
\begin{equation}
  L_{int} =      \sum_{\alpha \beta \gamma}\epsilon^{\alpha \beta \gamma} \sum_{i,j,k} G_{\psi,ijk}
  \bar{u}^{\alpha}_{i,R} (d^{\beta}_{j,R})^C \bar{d}^{\beta}_{k,R}\psi_L \,.
  \label{BARYONint}
\end{equation}
Here $\alpha, \beta, \gamma $ are colour indices,  $i, j, k$ are flavour
indices and $q_{L,R} = \frac{1 \pm \gamma_5}{2}q$.
The interaction (\ref{BARYONint}) is invariant under $SU_c(3) \otimes SU_2(L) \otimes U(1) $ the gauge interactions.
The simplest interaction of this type is
\begin{equation}
  L_{int} =   G_{\psi} \sum_{\alpha\beta\gamma}\epsilon^{\alpha \beta \gamma}  
  \bar{u}^{\alpha}_{R} (d^{\beta}_{R})^C \bar{s}^{\gamma}_{R}\psi_L \,.
  \label{BARYONint1}
\end{equation}
The interactions  (\ref{BARYONint},\ref{BARYONint1})
are  not hermitean  and they violate  the unitarity of the $S$-matrix. The interaction (\ref{BARYONint1})
describes  the reactions
\begin{equation}
  \bar{u} + \bar{d} \rightarrow s +  \bar{\psi} \,,
  \label{reac1}
\end{equation}
\begin{equation}
  \bar{d} + \bar{s} \rightarrow u +  \bar{\psi} \,,
  \label{reac2}
\end{equation}
\begin{equation}
  \bar{u} + \bar{s} \rightarrow d +  \bar{\psi} \,.
  \label{reac3}
\end{equation}
It is important that $C$-conjugate reactions
\begin{equation}
  u + d \rightarrow \bar{s} +  \psi    \,,
  \label{reac1a}
\end{equation}
\begin{equation}
  d +  s \rightarrow \bar{u} +  \psi \,,
  \label{reac2a}
\end{equation}
\begin{equation}
  u +  s \rightarrow \bar{d} +  \psi  \,.
  \label{reac2a}
\end{equation}
are prohibited for the interaction (\ref{BARYONint1}).
The reactions (\ref{reac1}, \ref{reac2}, \ref{reac3}) lead to the appearance of the baryon asymmetry of the Universe.
The $\psi$ field has baryon number equal to 1 and the
interactions        ( \ref{BARYONint}, \ref{BARYONint1}) conserve baryon number.   So in our scenario the $\bar{\psi}$ particles are
the dark matter particles. Moreover $\psi$-particles   have nonzero baryon number and the total baryon number of the
Universe being 
the sum of the baryon number of the matter and dark matter  is equal to zero.
  We assume that at the early Universe immidiately after inflation phase  the density of $\psi$ particles is negligible and
the Universe has zero baryon number, i.e. the number of quarks and antiquarks coincide.
The reactions (\ref{reac1}, \ref{reac2}) lead to the appearance of the baryon asymmetry of the Universe.
Antiquarks annihilate into $\bar{\psi}$ particle and quarks  while due to the specifics of the nonhermitean interaction (\ref{BARYONint})
quarks do not annihilate into $\psi$  particles.
To estimate the baryon asymmetry of the Universe we  use the Boltzman equation \cite{ParticleData}
\begin{equation}
  \frac{dn_{\Delta q}(t)}{dt} + 3 H(t)n_{\Delta q}(t) = <\sigma v_{rel}> n^2_q(t) \,,
  \label{Bolt}
  \end{equation}
Here $H(t)$ is the Hubble constant at time $t$, $n_q(t)$ is the density of quarks and   $n_{\Delta q}(t)$ is the density excess of quarks
over antiquarks $n_{\Delta q}(t) = n_q(t) - n _{\bar{q}}(t)$,   $  <\sigma v_{rel}>$ is the average annihilation
cross section.
In our estimates we take $H(T) = 1.66 N(T)^{1/2}\frac{T^2}{M_{PL}}$, 
$ t = (\frac{90}{32\pi^3 G_N N(T)^{1/2}})^{1/2} T^{-2}  $, $4N(T) = 427$.
We assume that
$n_{\Delta q}(T_h) = 0 $ at some high temperature $T_h$(reheating temperature).
Also we use the relation \cite{ParticleData} $\Delta n_B \approx 6 \times 10^{-10} n_{\gamma}$ between the baryon excess number and the number
of photons.  As an approximate solution of the Boltzman equation we
obtain the following estimate:
\begin{equation}
  G^2_{\psi}T^3_hM_{PL} \sim O(10^{-9}) \,.
  \label{estimate}
\end{equation}
For $G_{\psi} = k \cdot M^{-2}_{PL} $ we find $ T_h \sim  O( 10^{16}) \cdot k^{-2/3}~GeV$.
It should be stressed  that in the reaction
$ \bar{q} + \bar{q^{'}} \rightarrow q^{''} + \bar{\psi} $ we don't have baryon  number violation.
For the interactions (\ref{BARYONint}, \ref{BARYONint1})   $\bar{\psi}$ particles don't interact with the SM particles.
  We can identify the  $\bar{\psi}$ particles as dark matter particles. Experimental data give the ratio of dark matter density to baryon density
  of  the Universe $\frac{\rho_{DM}}{\rho_{B}} \approx 6$ \cite{ParticleData}. It means that in the considered scenario the mass of the $\bar{\psi}$
  particle is equal to $ \sim~6~GeV$.
So we find that nonhermitean interaction (\ref{BARYONint}) of antiquarks with new particle $\psi$ leads to the annihilation of antiquarks into $\bar{\psi}$
  particles without violation of baryon number. The key element here is that due to postulated nonhermitean interactions
  (\ref{BARYONint},\ref{BARYONint1})  quarks don't annihilate into $\psi$ particles. Therefore we have only the $\bar{\psi}$ particles  production
  in  the early Universe.  Moreover we can identify the $\bar{\psi}$ particles as dark matter particles with a mass $\sim 6~GeV$.
  The proposed model uses  nonrenormalizable interaction  ( \ref{BARYONint1}). 
 It is possible to construct renormalizable model explaining both the asymmetry of the Universe and dark matter.  
  Namely  we introduce new colour triplet scalar field $\phi_a(x)$. The field $\phi_a(x)$ has
  renormalizable $SU_c(3)\otimes SU_L(2) \otimes U(1)$ invariant nonhermitean  Yukawa interaction with quarks
  \begin{equation}
    L_{Yuk} =h_{1\phi}\phi_a(x)(  \bar{\psi}(x)  d_{Ra}(x))^*  + h_{2\phi}\phi^{*}_a(x)(\epsilon_{abc}u_{Rb}(x)d_{Rb}(x))^* \,.
    \label{model2}
  \end{equation}
  Here $a,b,c = 1,2,3$ are colour indices and $u_{Ra}d_{Rb}  = \epsilon_{ij}u_{Ra}^{i}
  d^{j}_{Rb} $  The field $\phi(x)$ has electroweak hypercharge $Y_{\phi} = - 1/3$ and the electric charge $Q_{\phi} = - 1/3$.
  The reaction
  \begin{equation}
    g + g \rightarrow \phi_a (\phi_{a})^* \rightarrow   \bar{\psi} d_{Ra}u_{Rb}u_{Rc} \,
    \label{reac1}
  \end{equation}
   leads to the baryon and dark matter production. Here $g$ is gluon. The field $\psi(x)$
    has zero hypercharge and nonzero baryon number  $B(\psi) = -1$. As in
    a previous model the mass of the $\psi$ particle  is predicted equal to $m_{\psi} \approx 6~GeV$.
    Note that dark matter particles interact with ordinary particles (electrons, nucleons, ...) mainly via
    gravitational interactions and the corresponding cross sections are extremely small.  As a
    consequence all underground experiments will fail to discover  dark matter.

\section{Conclusions}

In this paper we studied quantum field theories with unitarity violation. Quantum field theories with unitarity
violation correspond to quantum field theories with nonhermitean interaction Lagrangian. We constructed
several strictly solvable local field theory models with unitarity violation.
As an application of the considered models with unitarity violation 
we proposed two models  
   explaining observable  baryon asymmetry of the Universe and the existence of dark matter.
  In proposed models dark matter particles have nonzero  baryon number
    and
    the total baryon number of the Universe being the sum of the baryon number of the matter and dark matter  is equal to zero.  The mass of dark matter particles is predicted to be $\sim 6~GeV$.
  In considered models  dark matter particles interact with ordinary particles (electrons, nucleons, ...) mainly via
    gravitational interactions and the corresponding cross sections are extremely small.  As a
    consequence all underground experiments will fail to discover  dark matter. 
    Proposed models  also  violate   $CPT$ theorem,
  in particular, the decay width of the particle is  not equal to the decay width of the antiparticle. 
  The perspective way to look for  unitarity violation of the $S$-matrix is the study of the   $K$-meson decays.
  Namely it is very important to perform experimental check of  the CPT-predictions
  $\Gamma_{tot}(K^+) =\Gamma_{tot}(K^-)$ and  $\Gamma_{tot}(K^0) =\Gamma_{tot}(\bar{K}^0)$.



\section{ACKNOWLEDGMENTS}
I am indebted to collaborators of INR theoretical department for discussions. 





\newpage
\begin{thebibliography}{99}  
\bibitem{Bog} As a review see for example: \\
N.N.Bogoliubov and D.V.Shirkov, Introduction to the theory of quantized fields,  Chs. VIII, IX., Interscience, New York, (1959).
\bibitem{ParticleData} As a review see for example: \\
 PARTICLE PHYSICS BOOKLET, S.Navas et al., Phys.Rev. {\bf D110}, 030001 (2024).
 \bibitem{PAULI}   W.Pauli and F.Villars, Rev.Mod.Phys. {\bf 21}, 434-444 (1949). 
\bibitem{Slavnov} A.A.Slavnov, Theor.Math.Phys.  {\bf 13}, 174-177 (1977).
\bibitem{Stelle} K.S.Stelle,  Phys.Rev. {\bf D16}, 953-969 (1977).
\bibitem{Leon} J.H.Leon, C.P.Martin and F.Ruiz Ruiz,   
  Phys.Lett. {\bf B355}, 531 (1995).
\bibitem{Leon1} C.P.Martin and F.Ruiz Ruiz, , Nucl Phys.  {\bf B436}, 545 (1995).
\bibitem{Kubo} J.Kubo and T.Kugo,  PTEP2023, {\bf 12}, 123B02  (2023).
  \bibitem{Krasnikov} N.V.Krasnikov,  Theor.Math.Phys. {\bf 228:1}, 65 (2026);
arXiv:2510.14563.
\bibitem{Rubakov}
    Valery A.Rubakov,  and Dmitry Gorbunov, ``Introduction to the Theory of the Early Universe'', \\
 1st edn. (World Scientific Pub. Co., Singapore, 2017).
\bibitem{Kolb}
  Edward W.Kolb and Michael Turner,   ``The Early Universe'' \\
  FERMILAB-BOOK-1990-01, 1990.
\bibitem{B1} A.D.Sakharov, 
  JETP Lett., {\bf 5}, 24 (1967).
\bibitem{B2} A.Yu.Ignatiev, N.V.Krasnikov, V.A.Kuzmin and A.N.Tavkhelidze, Phys.Lett. {\bf 76  },  431 (1978).
\bibitem{B3} M.Yoshimura, Phys.Rev.Lett.  {\bf 41},  281 (1978).
\bibitem{B4} H.Davoudiasl, D.F.Morriseg, K.Sugurdson and S.Tulin, Phys.Rev.Lett. {\bf 105},  211304 (2010).
\bibitem{B5} As a review see for example: \\
V.A.Rubakov and M.E.Shaposhnikov, Phys.Usp. {\bf 39}, 461 (1996). 
 \bibitem{Zurek}  K.M.Zurek,  Phys.Rept. {\bf 537}, 9 (2014).

\end{thebibliography}
\end{document}